# Continuous metal-insulator transition of the antiferromagnetic perovskite NaOsO$_3$

Y.G. Shi,[1,2] Y.F. Guo,[1,2] S. Yu,[3] M. Arai,[4] A.A. Belik,[1,2] A. Sato,[5] K. Yamaura,[2,3,*]
E. Takayama-Muromachi,[1,2,3] H.F. Tian,[6] H.X. Yang,[6] J.Q. Li,[6]
T. Varga,[7] J.F. Mitchell,[7] S. Okamoto[8]

[1] International Center for Materials Nanoarchitectonics (MANA), [2] JST, Transformative Research-Project on Iron Pnictides (TRIP), [3] Superconducting Materials Center, [4] Computational Materials Science Center, [5] Materials Analysis Station, National Institute for Materials Science, 1-1 Namiki, Tsukuba, Ibaraki 305-0044, Japan

[6] Beijing National Laboratory for Condensed Matter Physics, Institute of Physics, Chinese Academy of Sciences, Beijing 100190, China

[7] Materials Science Division, Argonne National Laboratory, 9700 South Cass Avenue, Argonne, IL 60439, USA

[8] Materials Science and Technology Division, Oak Ridge National Laboratory, Oak Ridge, TN 37831, USA

**Abstract**

Newly synthesized perovskite NaOsO$_3$ shows Curie-Weiss metallic nature at high temperature and suddenly goes into an antiferromagnetically insulating state at 410 K on cooling. Electronic specific heat at the low temperature limit is absent, indicating that the band gap fully opens. *In situ* observation in electron microscopy undetected any lattice anomalies in the vicinity of the transition temperature. It is most likely that the antiferromagnetic correlation plays an essential role of the gap opening.

* E-mail address: YAMAURA.Kazunari@nims.go.jp

Metal-insulator transition (MIT) is a central topic of condensed matter science for more than half a century, and it still attracts intense attentions because it underlies key principles of correlated electrons science. Chronologically, a principal idea was proposed by Mott in 40s and Hubbard in 60s, suggesting that strong coulomb interaction in half-filling opens a gap at the Fermi energy ($E_F$), turning the conducting state into an insulating state, regardless of presence of magnetic correlations [1,2]. Alternatively, Slater suggested in 50s that antiferromagnetic (AF) order alone can open a gap in the half-filled state regardless of magnitude of the coulomb interaction [3]. Experimentally, many numbers of conductors were found to show MIT which seems to be well characterized in the above schemes [4]. Regarding the AF correlation-induced MIT, most discoveries were attained on low dimensional conductors, and fairly few were on 3-dimensional (3D) conductors. Generally, a 3D conductor causes only a small fraction of changes in its electronic structure through the AF ordering, resulting in robust conductivity such as found in Cr [5]. In this context, the 3D conductors $Cd_2Os_2O_7$ [6,7] and $Ln_2Ir_2O_7$ [8] are outstanding because of those continuous AF MITs.

So-far studies on $Cd_2Os_2O_7$ suggested that the continuous MIT at 226 K can be characterized in terms of the AF correlation [6,7], although inherent magnetic frustration due to the pyrochlore lattice certainly complicates the MIT. The origin of the MIT of $Ln_2Ir_2O_7$ might include the magnetic frustration somewhat [8]. It appears that the AF 3D MIT needs to be studied further. It is thus highly desirable to investigate an AF 3D MIT that is less relevant to the magnetic frustration.

We report a newly synthesized perovskite showing a continuous MIT. The perovskite $NaOsO_3$ has the octahedral environment of $Os^{5+}O_6$ so that the electronic configuration is $5d^3$, suggesting the $t_{2g}$ band is nearly half filling. $NaOsO_3$ shows Curie-Weiss metallic nature and abruptly turns to an AF insulator at 410 K on cooling. It is notable that an AF 3D MIT appears on the perovskite lattice rather than the pyrochlore lattice. Because $NaOsO_3$ is expected to be less relevant to the magnetic frustration unlike the isoelectronic $Cd_2Os_2O_7$ and charge ordering is absent over the MIT of $NaOsO_3$, the newly synthesized perovskite is therefore significant to provide valuable opportunities to deepen our understanding of AF 3D MIT.

Single crystal of $NaOsO_3$ grew up in a high-pressure apparatus which is capable of maintain 6 GPa during heating at 1700 °C for 2.5 hrs. The starting materials were $Na_2O_2$ (97%, Sigma-Aldrich) and $OsO_2$ (Os-84.0%, Alfa Aesar), and those were sealed in a Pt capsule at 15 mole % Na-rich stoichiometry with 0.1 mole of NaCl (99.99%, Rare metallic Co.) per the formula unit. After the heating, the capsule was quenched in the press to ambient temperature before releasing the pressure. We prepared a polycrystalline sample as well by heating at 1200 °C for 1 hr in the press without using NaCl. A photo of a selected crystal is shown in Fig. 1a. The samples were rinsed in water in a sonic bath for 2-3 minutes several times to remove residual ingredients, followed by drying in air at 140 °C for 10 min. Sample quality was checked by a powder X-ray diffraction (XRD) method using CuK$\alpha$ radiation in RINT 2200V, RIGAKU, confirming absence of impurities. We should note that preliminary samples prepared without excess $Na_2O_2$ were found to contain non-trivial amount of $OsO_2$. The crystal structure of $NaOsO_3$ was studied by a Rietveld method with powder synchrotron XRD data collected on the X-ray Operations and Research beamline at the Advanced Photon Source, Argonne National Laboratory. The incident beam was monochromatized at $\lambda = 0.401036$ Å. The Rietveld analysis was carried out by the program RIETAN-2000 [9].

A selected crystal of $NaOsO_3$ was examined in electron probe micro-analysis (JXA-8500F, JEOL), finding absence of contaminations such as Pt. The mean metal ratio at 5 points in the sole crystal was Na/Os = 1.14(2), suggesting small amount of Na is possibly incorporated in the crystal. Crystals were studied by a selected area electron diffraction (SAED) method between room temperature and 600 K in a transmission electron microscope operated at 200 kV (Tecnai-F20, Philips Electron Optics). Electrical resistivity ($\rho$) of a selected crystal was measured by a van der Pauw method on the assumption that the charge transport of the crystal is isotropic in nature with a dc-gage current of 0.1 mA between 2 K and 330 K in Physical Properties Measurement System (PPMS), Quantum Design. Electrical contacts on the 4 corners of the crystal were prepared by gold wires and silver paste. The limited size of the crystal did not allow us to estimate actual degree of the charge transport anisotropy. $\rho$ measurements between 300 K and 550 K were conducted in a laboratory-made apparatus. Hall coefficient ($R_H$) was measured by rotating the crystal by 180 degrees in a magnetic field of 50 kOe in PPMS between 25 K and 400 K.

Specific heat ($C_p$) was measured by a time-relaxation method using the powder sample (compressed) in PPMS between 2 K and 300 K. Differential scanning calorimetry (DSC) was conducted in DSC1 STAR$^e$ System, Mettler Toledo, between 300 K and 440 K at heating rations of 0.5, 2, 5, 10, and 20 K/min. Magnetic susceptibility ($\chi$) and isothermal magnetization were measured using multiple single crystals randomly oriented in a sample holder (~20 pieces, 8.1 mg in total) in Magnetic Properties Measurement System, Quantum Design.   The sample was cooled to 2 K without applying a magnetic field, and then warmed up to 300 K in a field of 50 kOe (zero-field cooling, ZFC), followed by cooling to 2 K in the field (field cooling, FC).   The measurement was again conducted between 300 K and 600 K in an oven installed to the magnetometer.

Fig. 1b shows the synchrotron XRD pattern with the analysis result [10].   A structure model with the space group *Pnma,* which often observed for perovskite oxides, was found to reasonably fit to the pattern.   The *R* factors and the difference curve in Fig. 1b indicate high-quality of the solution. Overall structure view is drawn in the Fig. 1b.   Based on the result, we investigated the local coordination of Os.   We found it is fairly isotropic: variation of the 6 bond-distances between Os and O is smaller than 0.4 % of the longest 1.946(1) Å bond, and the O-Os-O angles are 90.7, 89.9, and 90.1 degrees, being fairly close to the right angle.   Besides, in order to alternatively confirm the solution quality, bond valence sum of the atoms in the cell was calculated [11,12]: 4.92(Os) in reasonable agreement with the formal valence of Os and 1.40(Na) which is fair were obtained.

We also studied the perovskite lattice by a SAED method, and found all diffraction spots at room temperature are distributed in accord with the *Pnma* model, in short $\sqrt{2}\,a_\mathrm{p} \times 2a_\mathrm{p} \times \sqrt{2}\,a_\mathrm{p}$ ($a_\mathrm{p}$: primitive cell constant) order was clearly confirmed.   The superstructure is due to the cooperative rotation/tilting of the $OsO_6$ octahedra as in the $GdFeO_3$-type perovskite [13].   Additional superstructure neither commensurate nor incommensurate was detected at room temperature.   A selected SAED pattern is shown in Fig. 1c.   Besides, we carried out *in situ* heating in the microscope to test appearance of a possible lattice anomaly over the MIT.   However, no visible changes were detected to 600 K, excluding a major structure change and strong lattice-electrons coupling in the vicinity of the MIT temperature.   However, a possibility of structural distortions in local without symmetry change still remains, being investigated in future studies.

Fig. 2a shows temperature dependence of $\rho$ of the single crystal $NaOsO_3$.   The higher temperature part shows a slightly positive slope around the resistivity of $\sim 1\text{-}2 \times 10^{-4}$ $\Omega$cm, being consistent with an expected character for a metallic oxide.   The $\rho$ suddenly jumps up at 410 K on cooling and continuously rises to 2 K over 6 orders of magnitude.   It is notable that the warming and the cooling curves follow the same trace within an experimental accuracy, suggesting that the MIT is the second-order phase transition. The DSC confirmed the order of the transition (see inset to Fig.2), as the heat-flow-peak temperatures in cooling and heating are identical. The DSC measurements were repeated at different rates to extrapolate the result for 0 K/min, finding no shifts in the peak onsets and the peak positions.

In Fig. 2b, the Hall number ($e/R_H$) steadily decreases with temperature decreasing beyond 5 orders of magnitude, indicating that the carrier density is reduced continuously by the gap opening. The carrier density measured at 400 K (the technical limit) corresponds to 1.31 holes per the primitive cell, indicating that $NaOsO_3$ has sufficient amount of positively charged carriers. The estimated Hall mobility at 400 K is 1.3 cm$^2$ V$^{-1}$ s$^{-1}$, being not far from that for $Cd_2Os_2O_7$ [6]. We also estimated the Hall mobility at 200 K and 25 K; it is 13.8 cm$^2$ V$^{-1}$ s$^{-1}$ and 1.2 cm$^2$ V$^{-1}$ s$^{-1}$, respectively. It appears that the Hall-mobility changes one magnitude; however the change is too small to account for the large ρ change beyond 6 orders of magnitude. We thus conclude that the MIT is not due to losing the charge mobility.

We measured temperature dependence of $C_p$ of $NaOsO_3$ in order to further study the MIT: $C_p$ vs. $T$ is shown in Fig. 2c. From 2 K to 300 K, $C_p$ changes rather monotonically without manifest anomalies, suggesting absence of additional transitions. To parameterize the $C_p$, the curve was analyzed by the Debye model. The fit was conducted by a least-squares method and the best result (shown as a broken curve) yielded Debye temperature ($T_D$) of 505(5) K and the number of vibrating modes per formula unit in the Debye model ($n_D$) of 0.882(6)×5. Besides, the low-temperature part of $C_p$ was analyzed using the approximated Debye model $C(T)/T = \beta T^2 + \gamma$, where β and γ is a coefficient and the electronic-specific-heat coefficient, respectively (see the inset to Fig. 2c). In accord with the model, the data changes linearly (<14 K), yielding β of 9.83(5)×10$^{-5}$ J mol$^{-1}$ K$^{-4}$ and γ = 0.00(5) mJ mol$^{-1}$ K$^{-2}$. We obtained $T_D$ of 462.4(8) K from the β, where $n_D$ was assumed 5. It appeared that the γ is practically absent, suggesting that the electronic state is fully gaped at the low-temperature limit. For a comparison, γ estimated from the density of states (DOS without $U$, discussed later) is 7 mJ mol$^{-1}$ K$^{-2}$, being distinguishable from the observed γ.

Fig. 3a shows the temperature dependence of χ. First of all we should state that the discontinuous gap in the FC curve at 300 K is solely due to a technical matter regarding the heating attachment and does not entirely reflect the magnetism of the sample. Second, it is clear that χ vs. $T$

curves show a prominent anomaly at 410 K in the ZFC/FC conditions, indicating an establishment of a long-range magnetic order. We found that the magnetic transition is coincident with the MIT ($\rho$ vs. $T$). The isothermal magnetization in Fig. 3b displays evolution of the magnetization over the transition. We found that the spontaneous magnetization is fairly small, 0.005 $\mu_B$/Os at 5 K, this corresponds to only 0.17 % of the expected moment for $S = 3/2$, indicating that the transition is weakly ferromagnetic (FM).

In order to further analyze the magnetic properties, we applied the Curie-Weiss (CW) law to the paramagnetic part (>410 K), as shown in the inset to Fig. 3a. The analytical formula was $\chi(T) = N_A\mu_{eff}^2/3\pi(T-\Theta_W)$, where $N_A$ is the Avogadro constant and $\Theta_W$ is the Weiss temperature. The effective Bohr magneton ($\mu_{eff}$) was estimated to be 2.71 $\mu_B$, being 70 % of the expected moment for $S = 3/2$. The $\Theta_W$ was -1949 K, suggesting AF interactions are strong and dominant in the spin system if the CW analysis provides a valid indication. The overall magnetic data including the CW analysis results thus suggest that NaOsO$_3$ undergoes an AF transition at 410 K accompanying the weak magnetization most likely due to the Dzyaloshinsky-Moriya interaction generated by the broken inversion symmetry.

We studied the electronic structure of NaOsO$_3$ using the local density approximation (LDA) [14] of density functional theory [15]. We used the WIEN2k package [16], which is based on a full-potential augmented-plane-wave method. Experimental lattice parameters and atomic coordinates were used with the atomic radii of 2.0, 1.8, and 1.7 a.u. for Na, Os, and O, respectively. The cut-off wave-number $K$ in interstitial region was set to $RK = 7$, where $R$ is the smallest atomic radius. The integration over Brillouin zone was performed by a tetrahedron method with 144 k-points in the irreducible Brillouin zone (IBZ). For spin-polarized calculations, the convergence was checked with finer integration mesh points up to 468 k-points in the IBZ.

The DOS is plotted in Fig. 4a. The Os $t_{2g}$-bands distribute between -2.5 to 1 eV, being wider than those for Cd$_2$Os$_2$O$_7$ [17] because frustration is absent in NaOsO$_3$. Reflecting the nominal configuration of Os$^{5+}$ with 5$d^3$, the calculated $E_F$ lies near the center of Os $t_{2g}$-bands and total DOS shows the broad peak around the $E_F$. Since the half-filled $t_{2g}$-bands suggest instability of the non-magnetic phase, we examined possible magnetic solutions within LDA. However, any stable solutions with FM or AF spin alignments were found. Besides, we further performed LDA calculations with spin-orbit (SO) coupling since they may affect qualitatively the electronic structure, as intensively discussed for the 5$d$ oxide Sr$_2$IrO$_4$ [18]. Fig. 4b shows the total DOS with SO coupling. As is common with Cd$_2$Os$_2$O$_7$ [19], the SO coupling indeed modifies the $t_{2g}$-bands structure. However, the $E_F$ still locates at the vicinity of the broad peak with almost comparable DOS, suggesting the SO coupling plays an insignificant role of the gap opening in NaOsO$_3$ unlike what was found for Sr$_2$IrO$_4$ [20]. Stable magnetic solutions with the SO coupling were unattained.

Afterwards, we somehow obtained a stable magnetic solution by introducing a local coulomb interaction $U$ as the Hubbard term in the LDA+U approximation. At a moderate $U$ of 1 eV, a stable AF solution appeared with a small but finite energy gap as shown in Fig. 4c, while attempts with FM spin alignments never resulted in a corresponding solution qualitatively and quantitatively. Thus, the results suggest that the AF correlation is highly significant to open the gap in NaOsO$_3$. It is also suggested that a picture with a large on-site $U$ alone ($U \gg W$, where $W$ is the band width [1]) is too simple to account for the MIT of NaOsO$_3$.

The MIT of NaOsO$_3$ seems to be characterized by degree of magnitude of AF correlation, as discussed in Ref. 19. The theoretical study using dynamical mean field technique suggested that a gaped state in a half-filled electronic structure appears at the strong limit of the AF correlations, while a Mott transition is caused by strong coulomb repulsion regardless of magnitude of magnetic correlations. In fact, magnetically weak materials such as V$_2$O$_3$ and much magnetic materials such as NiS$_{2-x}$Se$_x$ show different MIT features [4]. Regarding NaOsO$_3$, the MIT coincides with the AF transition, NaOsO$_3$ is thus close to the strong magnetic limit of the 3D MIT [19]. We believe that such the paramagnetic metal to AF insulator transition at finite temperature is of 2nd order as we observed. Further discussion is left for future work.

On the other hand, a charge order model was intensively discussed for the 3$d$ perovskite $Ln$NiO$_3$ [20-21]: the MIT features are however readily distinguishable from what we found for NaOsO$_3$. For instance, SmNiO$_3$ shows a large temperature gap between the MIT at 400 K and a subsequent magnetic transition at 220 K [22]. In addition, the magnetic transition in SmNiO$_3$ is coupled with lattice alteration probably due to ordering between the $t_{2g}$ and $e_g$ orbitals, while such the strong electrons-lattice coupling is absent in NaOsO$_3$. It is thus reasonable to exclude the charge-order model from framework of the MIT of NaOsO$_3$.

In summary, the newly synthesized perovskite NaOsO$_3$ by means of a high-pressure method shows a dramatic MIT at 410 K. The MIT nature is far different from what we observed for $Ln$NiO$_3$ since it is associated with the magnetic ordering and is less relevant to the lattice anomaly. Furthermore, it is possibly different from the MIT nature of the isoelectrical Cd$_2$Os$_2$O$_7$ because the degree of magnetic frustration is highly reduced and the SO coupling is insignificant. The first-principles calculation suggests that the intense AF correlation (with small $U$) is the principal origin of the MIT of NaOsO$_3$. If the picture is true, NaOsO$_3$ is a "Slater insulator" in 3D, which can work at room temperature. Further studies toward nature of the MIT and possible practical applications are in progress.

We thank Dr. D. Mandrus for valuable discussion, Dr. T. Kolodiazhnyi for the $R_H$ measurement, and K. Kosuda for the EPMA. Use of the Advanced Photon Source was supported by the U.S. Department of Energy, Office of Science, Office of Basic Energy Sciences, under Contract No. DE-AC02-06CH11357. This research was supported in part by the WPI Initiative on Materials Nanoarchitectonics from MEXT, Japan, and the Grants-in-Aid for Scientific Research (20360012) from JSPS.

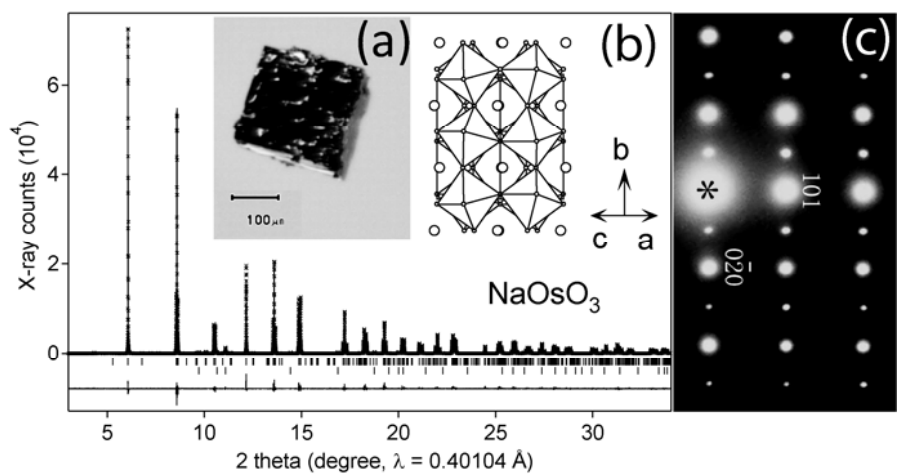

Fig. 1  (a) Photo of a crystal of NaOsO$_3$, and (b) the Rietveld analysis of the synchrotron XRD profile for NaOsO$_3$.  Cross markers and solid lines show the observed and calculated XRD profiles, respectively, and the difference is shown at the bottom.  The positions of Bragg reflections are marked by ticks (lower ticks are for the impurity OsO$_2$).  Inset is a structure view of NaOsO$_3$.  (c) SAED pattern taken along [10-1] zone axis at room temperature.

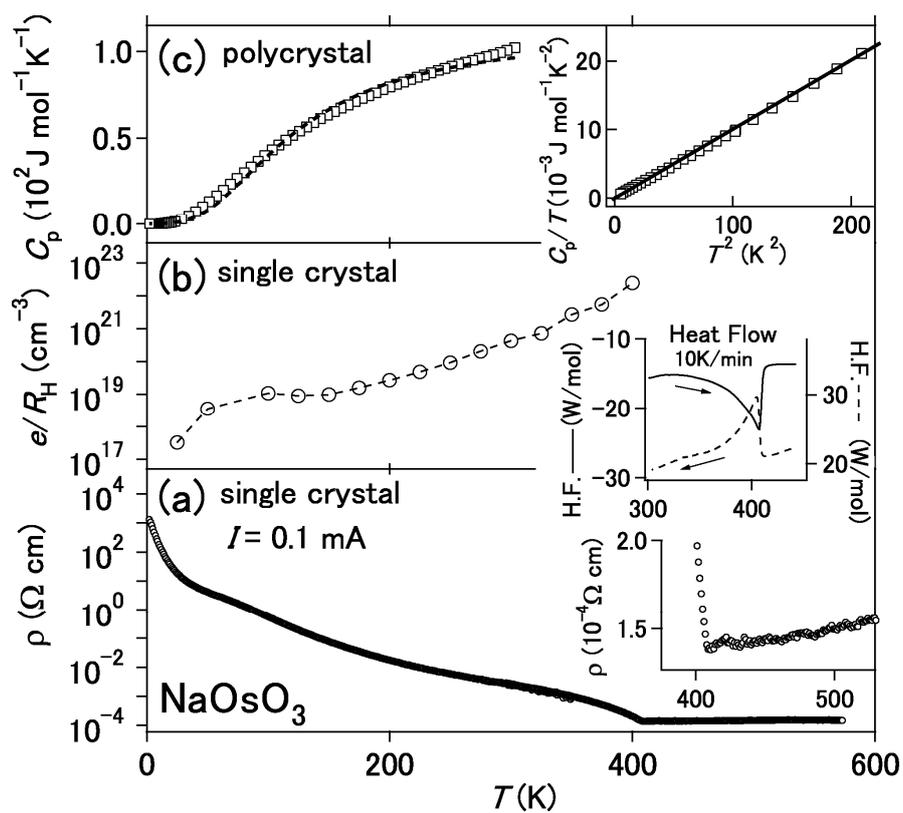

Fig. 2  $T$ dependence of (a) $\rho$, (b) $R_H$, and (c) $C_p$ of NaOsO$_3$.  Inset shows an expansion of $\rho$ vs. $T$, DSC curves, and $C_p/T$ vs. $T^2$ (solid line is a fit to the data) at the low temperature limit.  The data were measured by a van der Pauw method on the assumption that the charge transport of the crystal is isotropic in nature.

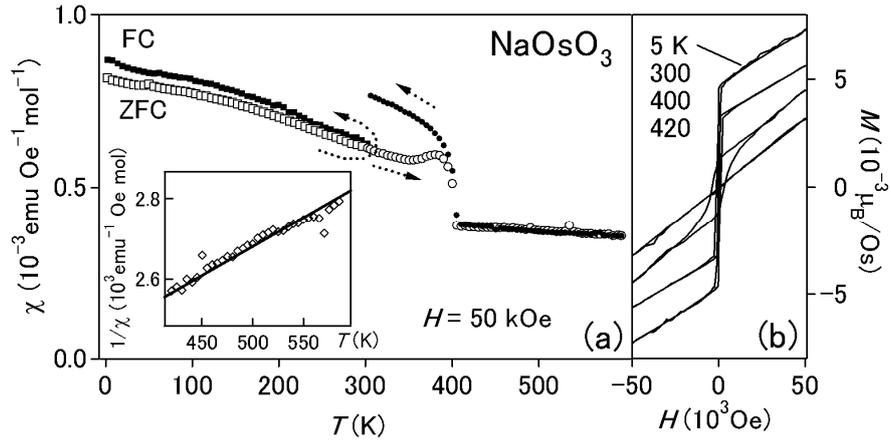

Fig. 3 (a) *T* dependence of χ of amount of single crystals of NaOsO$_3$, measured separately below (squares) and above 300 K (circles). Inset shows an alternative plot of the data above 410 K. (b) Isothermal magnetization of NaOsO$_3$. The measurements were conducted on crystals randomly oriented in a sample holder.

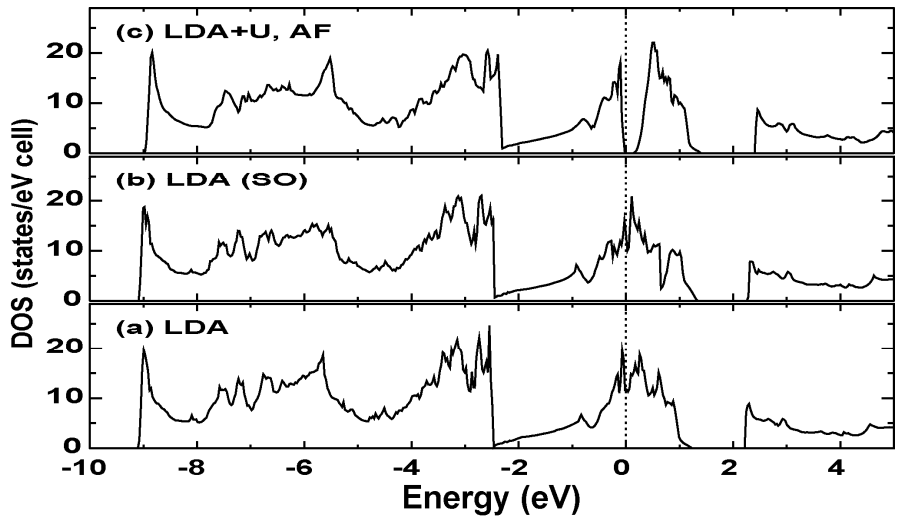

Fig. 4 (a) Non-magnetic DOS of NaOsO$_3$ (b) with SO coupling. (c) AF DOS with U.

13